\documentclass[conference]{IEEEtran}
\ifCLASSINFOpdf
\else
\fi

 \usepackage{graphicx}

\usepackage{flushend}

\begin{document}
%
\title{Multi-User Multi-Carrier  Differential Chaos Shift Keying Communication System}
%
%
%

\author{\IEEEauthorblockN{Georges Kaddoum, Fran\c cois-Dominique Richardson, Sarra Adouni, Fran\c cois Gagnon, Claude Thibeault}
\IEEEauthorblockA{Department of electrical engineer, LACIME Laboratory$^{*}$ \\ 
Universit\'{e} du Qu\'{e}bec, \'{E}cole de technologie sup\'{e}rieure \\
  Montreal,  Canada \\
E-mail:  georges.kaddoum@lacime.etsmtl.ca, francois.richardson@lacime.etsmtl.ca, ado.sa@live.fr \\
 francois.gagnon@etsmtl.ca, claude.thibeault@etsmtl.ca}
}


\IEEEoverridecommandlockouts
\IEEEpubid{\makebox[\columnwidth]{978-1-4673-2480-9/13/\$31.00 ~\copyright~2013 IEEE \hfill} \hspace{\columnsep}\makebox[\columnwidth]{ }}

\maketitle

{\let\thefootnote\relax\footnotetext{* This work has been supported in part by Ultra Electronics TCS and the Natural Science and Engineering Council of Canada as part of the 'High Performance Emergency and Tactical Wireless Communication Chair' at \'{E}cole de technologie sup\'{e}rieure.}}

\begin{abstract}
In this paper, a multi user Multi-Carrier Differential Chaos Shift Keying (MC-DCSK) modulation is presented. The system endeavors to provide a good trade-off between robustness, energy efficiency and high data rate, while still being simple. In this architecture of MC-DCSK system, for each user, chaotic reference sequence is transmitted over a predefined subcarrier frequency. Multiple modulated data streams are transmitted over the remaining subcarriers allocated for each user. This transmitter structure saves energy and increases the spectral efficiency of the conventional DCSK system.  
\end{abstract}




%
\IEEEpeerreviewmaketitle

\section{Introduction}

The demand for wireless services is in constant rise. Multi-carrier (MC) transmission, since it has the advantages of high spectral efficiency, robustness to frequency selective fading, and feasibility of low-cost transceiver implementation is a strong candidate for many wireless applications. Several combinations of multi-carrier and Code Division Multiple Access (CDMA), are proposed in the literature~\cite{Hanzo03,  kond96}. In MC-CDMA, one-bit chips are spread over $M$ subcarriers in the frequency domain~\cite{Hanzo03}, while for MC-DS-CDMA, time and frequency spreading is used (\textit{i.e.} TF-domain spreading)~\cite{kond96}. 

The chaotic signals have been shown to be well suited for spread-spectrum modulation because of their inherent wideband characteristic~\cite{Lau03} \cite{Kad09} \cite{Kaddoum2013}, mitigation of fading channels, jamming resistance and  low probability of intercept (LPI) ~\cite{Yu2005}. In addition, chaos-based sequences give good results as compared to Gold and independent and identically distributed sequences for reducing the peak-to-average power ratio (PAPR)~\cite{Vitali2006}.  

A proposed system with a non-coherent receiver, named differential chaos shift keying (DCSK) system, in which chaotic synchronization is not used on the receiver side, delivers a good performance in multipath channels~\cite{Kol98a}. Furthermore, differential non-coherent systems are better suited than coherent ones for time and frequency selective channels~\cite{LeSaux2005}. In the DCSK system, each bit duration is divided into two equal slots. In the first slot, a reference chaotic signal is sent. Depending on the bit being sent, the reference signal is either repeated or multiplied by the factor $-1$ and transmitted in the second slot. A significant drawback of DCSK is that for each bit one reference and half the bit duration is spent sending non-information-bearing reference samples~\cite{Lau03}. This can be accounted as being energy-inefficient and a serious data rate reducer. In \cite{Hua12}, the spectral efficiency of the DCSK is improved, but the system receiver requires an RF delay line, which is not easy to implement because of the wide bandwidth involved. In a study to overcome the problem of RF delay in DCSK systems, Xu \textit{et al.} proposed a Code Shifted Differential Chaos Shift Keying (CS-DCSK) system~\cite{Xu11}. In their system, the reference and the information bearing signals are separated by Walsh code sequences, and then transmitted in the same time slot. For such systems, there is no need for a delay line at the receiver end. An improved version of the high spectral efficiency DCSK system by \cite{Xu11} is presented in \cite{kad12}, where chaotic codes are used instead of Walsh codes, with different receiver structures. Another design based on an ultra-wideband system using chaotic signals for low complexity, low cost, low power, and low rate is presented in \cite{chi08}.  

In this paper, we first introduce a new design of a multi-user, multi-carrier DCSK system (MC-DCSK). On the transmitter side, $M$ subcarriers are assigned for each user, where one subcarrier is used to transmitting the references slot, while the $M-1$ other frequencies will carry the transmitted bits.  The proposed system solves the RF delay line problem mentioned in ~\cite{Xu11}, provides from the properties of DCSK system in terms of resistance to interference, increases the data rate, and optimizes the transmitted energy of the DCSK system with a simple transmitter/receiver design. The analytical performance derivation of DCSK communication system is studied in \cite{kad12wpc} \cite{Fan13} \cite{Fan13i} \cite{Kad11iscas} and the transmission security is improved in \cite{KadFDR2012a}. In this paper, for the space available, we concentrate our efforts to explain the proposed system design, where the analytical derivation and cognitive multiple access techniques of MC-DCSK system will be studied in future work. 

In this paper, section~\ref{sec:Chaos} describes the characteristics of chaos-based systems with an emphasis on DCSK. Section  ~\ref{sec:SystemArchitecture} covers the architecture of the multi-user MC-DCSK system. The energy of the system is examined in section~\ref{EnergyEfficiency}. Simulation results and discussions are presented in section~\ref{sec:Simulation}, and concluding remarks are presented in section~\ref{sec:Conclusion}.

\begin{figure*}[htb!]
\centering
\includegraphics[width=15 cm]{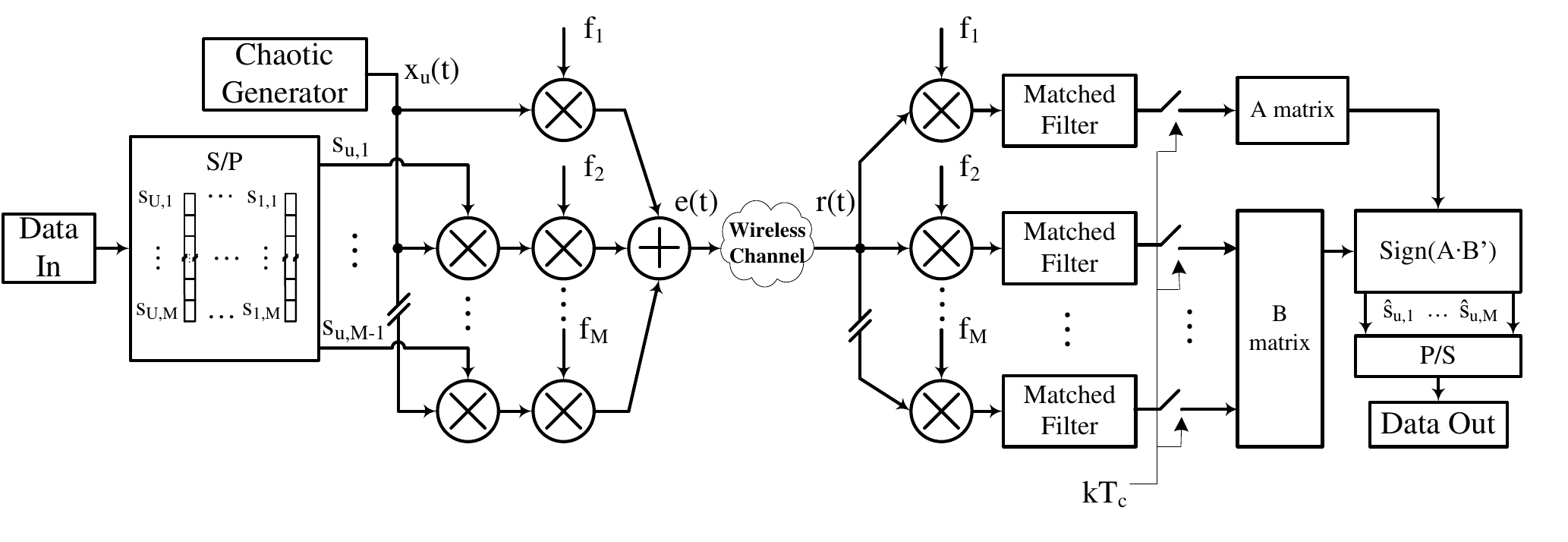}
\caption{Block diagram of the single-user MC-DCSK system \label{fig:TxRx}}

\end{figure*}

\section{Drawback of DCSK communication scheme}
\label{sec:Chaos}
In DCSK modulator, each bit $s_i~=~\lbrace -1, \: +1 \rbrace$ is represented by two sets of chaotic signal samples, with the first set representing the reference, and the second carrying data. If $+1$ is transmitted, the data-bearing sequence is equal to the reference sequence, and if $-1$ is transmitted, an inverted version of the reference sequence is used as the data-bearing sequence. 
Let $2\beta$ be the spreading factor, defined as the number of chaotic samples sent for each bit, where $\beta$ is an integer. 
During the $i^{th}$ bit duration, the output of the transmitter $e_{i,k}$ is
\begin{equation}\label{et}
e_{i,k} = \left\{ \begin{array}{l}
x_{i,k} \,\,\,\,\,\,\,\,\,\,\,\,\,\,\,\,\,\,\,{\rm for }\,\,1 < k \le \beta ,\\ 
s_i x_{i,k - \beta \,\,\,} \,\,  \,\,{\rm for }\,\,\beta < k \le 2\beta , \\ 
\end{array} \right.
\end{equation}
\\
where $x_k$ is the chaotic sequence used as reference and $x_{k-\beta}$ is the delayed version of the reference sequence.
 

%
%


In this system, half the bit duration time is spent sending a non-information-bearing reference. Therefore, the data rate of this architecture is seriously reduced compared to other systems using the same bandwidth, leading to a loss of energy. The reference sequence dissipates half the energy of each bit.

\section{Multi-User Multi-carrier DCSK system}
\label{sec:SystemArchitecture}

The system's architecture is intended to be of low complexity thus simple to implement. Numerous extensions could be performed to this system for different performance optimizations. 

\subsection{The transmitter}
\label{sec:Tx}
The MC-DCSK system benefits from the non-coherent advantages of DCSK and the spectral efficiency of multi-carrier modulation. 
Here, we consider an MC-DCSK system using discrete chaotic sequence values for modulation and square-root-raised-cosine filter chip waveforms. For mathematical simplification, the equations describe the MC-DCSK  system for one user. As shown in Figure~\ref{fig:TxRx}, for each user, a reference chaotic code $x_u$ is generated to be used as a reference and spreading code. The input information sequence is first converted into $U$ parallel data sequences $s_u(t)$ for $u=1,\: 2,...U$. The independent data sequence $s_u (t)$ with equal probability value is $+1,$ or $-1$, where 
\begin{equation}
s_u (t) = \sum\limits_{i =  1 }^{ M-1} {s_{u,i}} (t)\cdot
\end{equation} 
After a serial-to-parallel conversion, the $u^{th}$ ${M-1}$ bits substream is spread due to multiplication in time with the same chaotic spreading code $x_u$. 
 \begin{equation}
 x_u (t) = \sum\limits_{k =  1  }^{ \beta } {x_{u,k} } h(t-kT_c),
 \end{equation}  
 \\
where $h(t)$ is the square-root-raised-cosine filter. This filter is band-limited  and is normalized to have unit energy. Let $H(f)=\textit{F} 
\left\{ {h(t)} \right\}$, where $\textit{F} $ denotes a Fourier transform. It is assumed that $H(f)$ is limited to $[-B_c/2, \: B_c/2]$ which satisfies the Nyquist criterion with a rolloff factor $\alpha$ ($0 \le \alpha  \le 1 $) where $B_c=(1+\alpha)/T_c$.

The $x_u$ modulates the  subcarrier assigned to transmit the reference signal, after which the data signals spread  by ${M-1}$ modulate the ${M-1}$ subcarriers. 
Therefore, the transmitted signal of the mono-user MC-DCSK is given by:

\begin{equation}
\begin{array}{l}
 e(t) =  {x_{u} (t)\cos (2\pi f_1 t + \phi _1 )}  +  \\ 
 \sum\limits_{i = 2}^M {  {s_{u,i}^{} (t)x_{u} (t)\cos (2\pi f_i t + \phi _i )} }  \\ 
 \end{array} ,
\end{equation}
\\
where $\phi _i $ represents the phase angle introduced in the carrier modulation process. In this paper, we normalize the transmitted energy in every subcarrier.
 
For the MC-DCSK, the frequency corresponding to the $i^th$ subcarrier is $f_{i}= f_{p} + i/T_c$, where $f_p$ is the fundamental carrier frequency. The minimum spacing between two adjacent subcarriers equals $\Delta=(1 +\alpha)/T_c$, which is a widely used assumption.

\begin{figure}[htb!]
\centering
\includegraphics[width=9 cm]{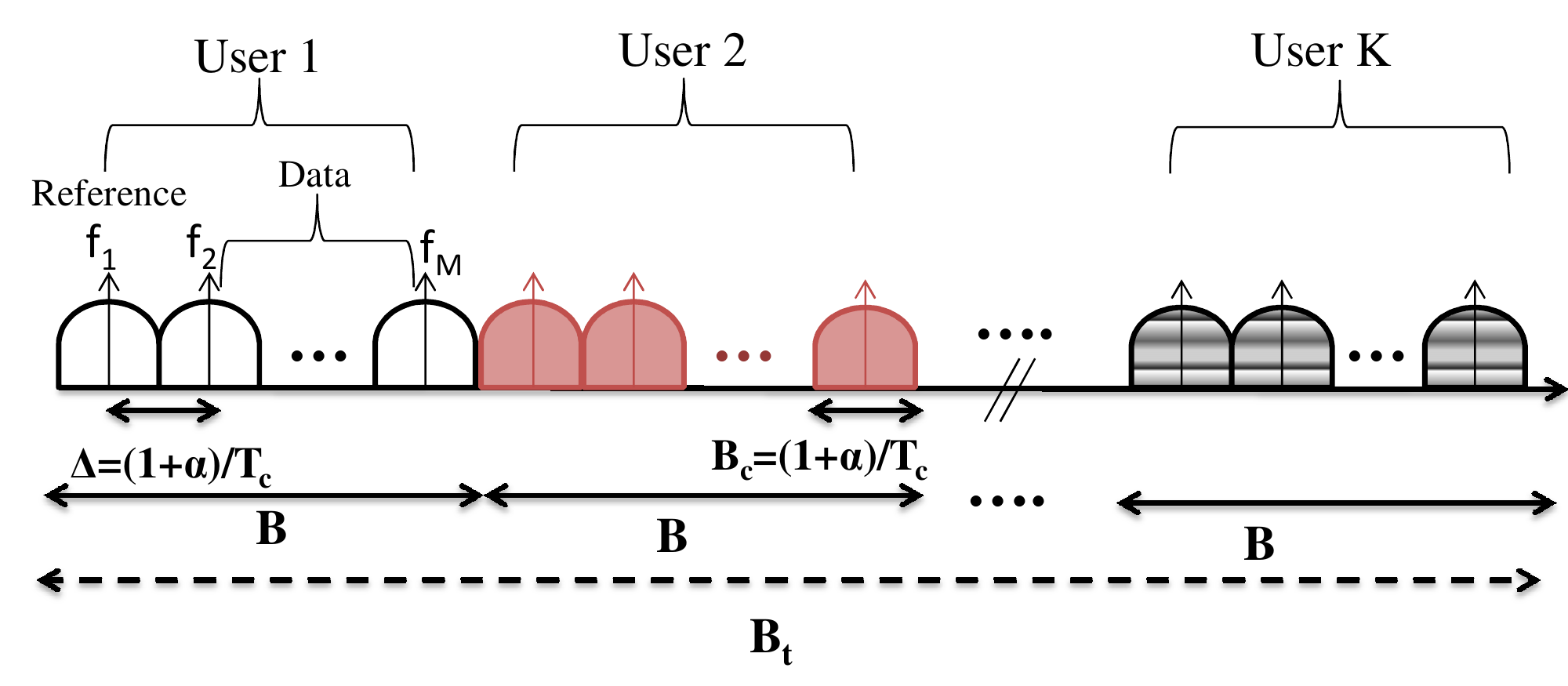}
\caption{The power spectral density of a band-limited multi-user MC-DCSK system. \label{psd_mc_DCSK}}

\end{figure}

Figure \ref{psd_mc_DCSK} shows the power spectral density (PSD) of the multi-user MC-DCSK system. Let $B$ be the bandwidth for each user. The maximal number of users using the channel in the same time is determined from the total allocated bandwidth $B_t$ for this system determine. When both $T_b$ and $B$ are set, the chip duration $T_c$ as well as the spreading factor $\beta$ depend on the number of carrier $M$, the bandwidth $B_c$ of each subchannel or the carrier spacing $\Delta$. In our design, we divide the total band $B$ into $M$ equi-width frequency bands, as shown in Figure \ref{psd_mc_DCSK}, where all bands are disjoint. The bandwidth of each carrier band $B_c$ is:

 \[ B_c= (1+ \alpha)/T_c ; \:  where \: B=M B_c ,\]

Thus, the spreading factor   $\beta = T_b/T_c $ function of the system parameters is :

\begin{equation} \label{sr_fc}
\beta  = \frac{{T_b B}}{{M(1 + \alpha )}} \cdot
\end{equation}

In order to validate our system design and evaluate its parameters, such as the spreading factor, the bit energy, and the bit error rate performance,  we assume that our channel is AWGN. In addition we assume that no interference between subcarriers. In this case we evaluate the received signal for one user.

\begin{equation}
r(t)= v(t)+n(t),
\end{equation}
\\
where $r(t)$ is the received signal, $n(t)$ is an AWGN noise with zero mean and power spectral density of $N_0/2$.

\subsection{The receiver}
\label{sec:Rx}

The block diagram of the MC-DCSK receiver is illustrated in Figure~\ref{fig:TxRx}. One of the objectives of this design was to provide a simple, easy-to-implement receiver providing good performance. We consider a set of matched filters, each demodulating the desired signal of the corresponding carrier frequency $f_i$, and then the signals are sampled every $kT_c$ time. The outputs discrete signals are stored in matrix memory. The matrix implementation of the receiver simplifies the parallel data recovery, where the decoding algorithm is: 

First, at the same time, the output of the first match is stored in matrix $P$ and the ${M-1}$ data signals are stored in the second matrix $S$, where:

\[P= (x_{u,1} + n_{u,1}, \: x_{u,2}+n_{u,2}, ... \: x_{u,\beta}+n_{u,\beta}) ,\] 
\\
where $n_{u,k}$ is the $k^{th}$  sample of additive Gaussian noise added to the reference signal.

The matrix $S$ is:

\[
S \hspace{-0.1 cm} = \hspace{-0.1cm} \left( \hspace{-0.2cm} {\begin{array}{*{20}c}
   {s_{u,1} x_{u,1} +n^{(1)}_{u,1} } &  \ldots  & {s_{u,1} x_{u,\beta } +n^{(1)}_{u,\beta}}  \\
    \vdots  &  \vdots  &  \vdots   \\
   {s_{u,M - 1} x_{u,1}  +n^{(M-1)}_{u,1} } &  \ldots  & {s_{u,M - 1} x_{u,\beta } +n^{(M-1)}_{u,\beta} }  \\ 
\end{array}} \hspace{-0.2cm} \right)\cdot
\]
\\
where $n^{(i)}_{u,k}$ is the $k^{th}$  sample of additive Gaussian noise added to the $i^{th}$ bit.

Finally, after $\beta$ clock cycles, all the samples are stored, and the decoding step is activated.
The transmitted ${M-1}$ bits are recovered in parallel by computing the sign of  the resultant vector of the matrix product:

\begin{equation}
\hat s_u  = sign(P \times S')\cdot
\end{equation}
\\
where $ \times $ is the matrix product and $'$ is the matrix transpose operator. In fact, this matrix product can be seen as a set of a parallel correlator where the reference signal multiplies each data slot, and the result is summed over the duration $\beta Tc$.


\section{Energy  efficiency }
\label{EnergyEfficiency}

The energy efficiency of the proposed system is improved as compared to the DCSK system. In fact, for the DCSK system, a new chaotic reference is generated for every transmitted bit, and in our case, one reference is shared with ${M-1}$ modulated bits. For a conventional DCSK system, the transmitted bit energy $E_b=E_{data}+E_{ref}$, where $E_{data}$  and $E_{ref}$ are the energies to transmit the data and reference respectively. 
Without loss of generality, the data and the reference energies are equal $  E_{data}=E_{ref}=T_c \sum\limits_{k = 1}^\beta  {x_{i,k}^2 }$. 
  
Then for DCSK system, the transmitted  energy for a given bit $i$ is $  E_b=2T_c \sum\limits_{k = 1}^\beta  {x_{i,k}^2 }$.

In the MC-DCSK system, the bit energy is the sum of the data carrier energy and a part of the reference energy:

\begin{equation}
E_b=E_{data}+ \frac{{E_{ref}}}{{M-1}} \cdot
\end{equation}

In our system, the energies on the $M$ subcarriers are equal:

\begin{equation}
E_{data}=E_{ref}=T_c \sum\limits_{k = 1}^\beta  {x_{i,k}^2 }.
 \end{equation}

To study the energy efficiency, we compute the transmitted Data-energy-to-Bit-energy Ratio (DBR):
\begin{equation}
\label{PowerRatio}
DBR=\frac{E_{data}}{E_{b}} \cdot
\end{equation}

In a conventional DCSK system, half the energy is transmitted into the reference to achieve modulation for each bit, and then the DBR is equal $
DBR=\frac{1}{2}$.

 In the MC-DCSK system the DBR is :

\begin{equation}
\label{PowerRatio}
DBR=\frac{M-1}{M} \cdot
\end{equation}

\begin{figure}[t]
\centering
\includegraphics[width=6.5 cm]{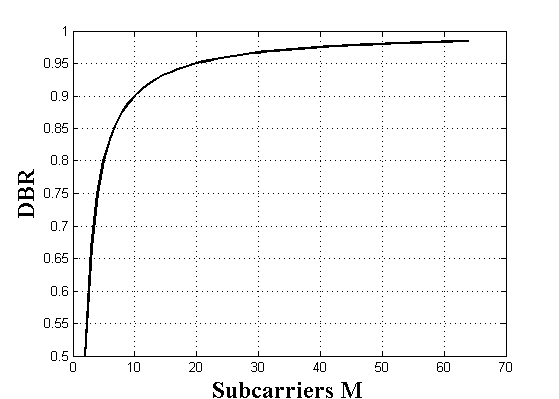}
\caption{DBR for a system for various amount of data subcarriers. \label{fig:PowerRatio}}

\end{figure}

As shown in Figure \ref{fig:PowerRatio}, for $M=2$ where we have one reference for one bit, in this case, the MC-DCSK system is equivalent to a DCSK system with $DBR=\frac{1}{2}$. This means that $50\%$ of the energy is used from the total bit energy to transmit the reference. We can see clearly that for $M > 20$, the reference energy accounts for less than $5\%$ of the total bit energy $E_b$.

\section{Simulation results and discussions}
\label{sec:Simulation}

Since the space between subcarriers guarantee a free interferences, the performance in mono-user or in multi user case in AWGN channel remains the same. 
The parameters of the simulation  are set as follows: the MC-DCSK system uses the square-root-raised-cosine chip waveform for a roll-off factor $\alpha$ equal to $0.25$. As shown in equation (\ref{sr_fc}), the spreading factor is computed as a function of the number of subcarriers $M$, the bit duration $T_b$, and the total allocated bandwidth $B$. In our simulations, we set the bit duration $T_b=400$, $B=1$ and for $M=64$ the allowed spreading factor $\beta =5$, for $M=16$ subcarriers $\beta =20$, for $M=8$ $\beta =40$, and for $M=2$ $\beta =160$.

\begin{figure}[htb!]
\centering
\includegraphics[width=7.2 cm]{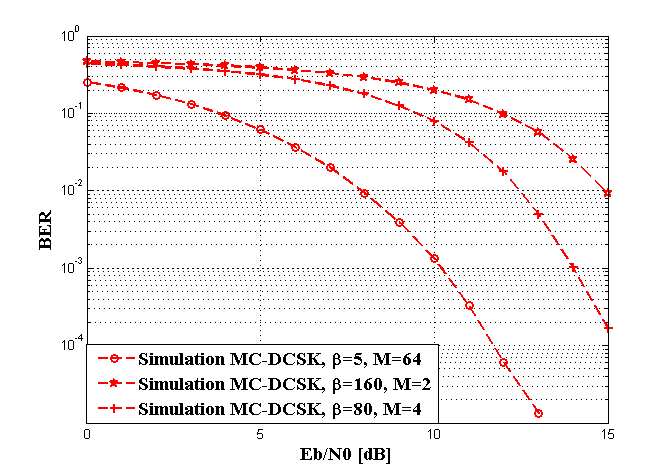}
\caption{BER performance of MC-DCSK for different number of subcarriers and spreading factors  \label{mc_DCSK_sf}}

\end{figure}

\begin{figure}[htb!]
\centering
\includegraphics[width=7.2 cm]{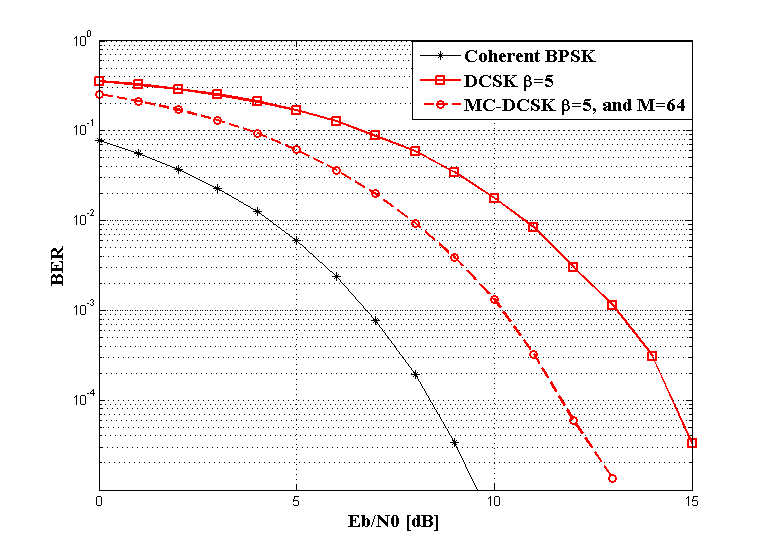}
\caption{BER comparison of MC-DCSK for M=64 and DCSK where the spreading factor $\beta=5$ \label{DCSK_mc_DCSK}}

\end{figure}

Figure \ref{mc_DCSK_sf} presents the performances obtained from the Monte Carlo simulations of the MC-DCSK system for different spreading factors and number of subcarriers $M$ per user. 
In Figure \ref{DCSK_mc_DCSK} we study the effect of the number of subcarriers on the system performance. To that end, we set the spreading factor to $\beta=5$ and the bit duration $T_b$, and then we assume that the bandwidth $B$ is wide enough to support any number of subcarriers $M$. Figure \ref{DCSK_mc_DCSK} shows  interesting results of our proposed MC-DCSK system in terms of performance enhancement. In fact, for a given spreading factor, when the number of subcarriers $M$ increases, the $DBR$ ratio tends toward one, meaning that less energy is used to transmit the reference code. This performance means that for high number of subcarriers $M$, we need less energy to reach a given BER. In this figure, we show the performance improvement by simulation for $M=2$ and $M=64$, with a fixed spreading factor equal to $\beta=5$. In this case, this result compares the performance of the proposed system with that of the conventional DCSK. In fact, when $M=2$, the MC-DCSK system is equivalent to a DCSK system because one reference is required to transmit every bit.  In the same figure, we can observe a degradation in performance between the MC-DCSK system for $M=64$ and the coherent BPSK one. This degradation comes from the two noise sources added to the reference and the data carrier signals.

\section{Conclusion}
\label{sec:Conclusion}
An energy-efficient non-coherent multi-carrier spread spectrum system has been presented. From the outstanding energy inefficiency drawback imposed by time-multiplexed differential modulations, a novel frequency multiplexed architecture is designed. The multi-carrier characteristic of this novel design enables significant energy savings and a higher spectral efficiency as compared to differential systems because for each user, the reference signal is only sent once for $M-1$ parallel bits. The energy efficiency of the proposed system is analyzed and a $DBR$ is derived, with results showing that for $M > 20$ subcarriers, the energy lost in transmitting the reference is less than $ 5\%$ of the total bit energy. To compare the performance of the proposed system with that of the DCSK, the simulated BERs are plotted with the same spreading factor, where results prove an increase in performance as compared to the conventional DCSK.  Our future work will focus on defining   new cognitive multi-user access strategies and performance improvement of this system.

\bibliographystyle{IEEEtran}
\bibliography{bibliographie}

\end{document}